\documentclass[10pt,a4paper,onecolumn,numbers]{elsarticle}
\usepackage[latin1]{inputenc}
\usepackage{amsmath}
\usepackage{amsfonts}
\usepackage{amssymb}
\usepackage{graphicx}
\usepackage{fancyhdr}
\usepackage{times}
\usepackage{natbib}

\begin{document}
\begin{frontmatter}
\title{\mbox{LOPES-3D} - vectorial measurements of radio emission from cosmic ray induced air showers}

\author[1]{W.D.~Apel}
\author[2,14]{J.C.~Arteaga}
\author[3]{L.~B\"ahren}
\author[1]{K.~Bekk}
\author[4]{M.~Bertaina}
\author[5]{P.L.~Biermann}
\author[1,2]{J.~Bl\"umer}
\author[1]{H.~Bozdog}
\author[6]{I.M.~Brancus}
\author[4]{A.~Chiavassa}
\author[1]{K.~Daumiller}
\author[2,15]{V.~de~Souza}
\author[4]{F.~Di~Pierro}
\author[1]{P.~Doll}
\author[1]{R.~Engel}
\author[3,9,5]{H.~Falcke}
\author[2]{B.~Fuchs}
\author[10]{D.~Fuhrmann}
\author[11]{H.~Gemmeke}
\author[7]{C.~Grupen}
\author[1]{A.~Haungs}
\author[1]{D.~Heck}
\author[3]{J.R.~H\"orandel}
\author[5]{A.~Horneffer}
\author[2]{D.~Huber\corref{cor}}
\ead{Daniel.Huber@kit.edu}
\author[1]{T.~Huege}
\author[1,16]{P.G.~Isar}
\author[10]{K.-H.~Kampert}
\author[2]{D.~Kang}
\author[11]{O.~Kr\"omer}
\author[3]{J.~Kuijpers}
\author[2]{K.~Link}
\author[12]{P.~{\L}uczak}
\author[2]{M.~Ludwig}
\author[1]{H.J.~Mathes}
\author[2]{M.~Melissas}
\author[8]{C.~Morello}
\author[1]{J.~Oehlschl\"ager}
\author[2]{N.~Palmieri}
\author[1]{T.~Pierog}
\author[10]{J.~Rautenberg}
\author[1]{H.~Rebel}
\author[1]{M.~Roth}
\author[11]{C.~R\"uhle}
\author[6]{A.~Saftoiu}
\author[1]{H.~Schieler}
\author[11]{A.~Schmidt}
\author[1]{F.G.~Schr\"oder}
\author[13]{O.~Sima}
\author[6]{G.~Toma}
\author[8]{G.C.~Trinchero}
\author[1]{A.~Weindl}
\author[1]{J.~Wochele}
\author[12]{J.~Zabierowski}
\author[5]{J.A.~Zensus}

\address[1]{Karlsruhe Institute of Technology (KIT), Institut f\"ur Kernphysik, Germany}
\address[2]{Karlsruhe Institute of Technology (KIT), Institut f\"ur Experimentelle Kernphysik, Germany}
\address[3]{Radboud University Nijmegen, Department of Astrophysics, The Netherlands}
\address[4]{Dipartimento di Fisica Generale dell' Universit\`a Torino, Italy}
\address[5]{Max-Planck-Institut f\"ur Radioastronomie Bonn, Germany}
\address[6]{National Institute of Physics and Nuclear Engineering, Bucharest, Romania}
\address[7]{Universit\"at Siegen, Fachbereich Physik, Germany}
\address[8]{INAF Torino, Instituto di Fisica dello Spazio Interplanetario, Italy}
\address[9]{ASTRON, Dwingeloo, The Netherlands}
\address[{10}]{Universit\"at Wuppertal, Fachbereich Physik, Germany}
\address[{11}]{Karlsruhe Institute of Technology (KIT), Institut f\"ur Prozessdatenverarbeitung und Elektronik, Germany}
\address[{12}]{National Centre for Nuclear Research, Department of Cosmic Ray Physics, {\L}\'{o}d\'{z}, Poland}
\address[{13}]{University of Bucharest, Department of Physics, Romania}
\scriptsize{
\address[{14}]{now at: Universidad Michoacana, Morelia, Mexico}
\address[{15}]{now at: Universidad S\~ao Paulo, Inst. de F\'{\i}sica de S\~ao Carlos, Brasil}
\address[{16}]{now at: Institute for Space Sciences, Bucharest, Romania}
}

\cortext[cor]{Corresponding author:}

\begin{abstract}
\mbox{LOPES-3D} is able to measure all three components of the electric field vector of the radio
emission from air showers. This allows a better comparison with emission models. The measurement of
the vertical component increases the sensitivity to inclined showers. By measuring all three components of
the electric field vector \mbox{LOPES-3D} demonstrates by how much the reconstruction accuracy of primary
cosmic ray parameters increases. Thus \mbox{LOPES-3D} evaluates the usefulness of vectorial measurements
for large scale applications.  
\end{abstract}

\end{frontmatter}

\begin{figure} \center
 \includegraphics[width=0.5 \columnwidth]{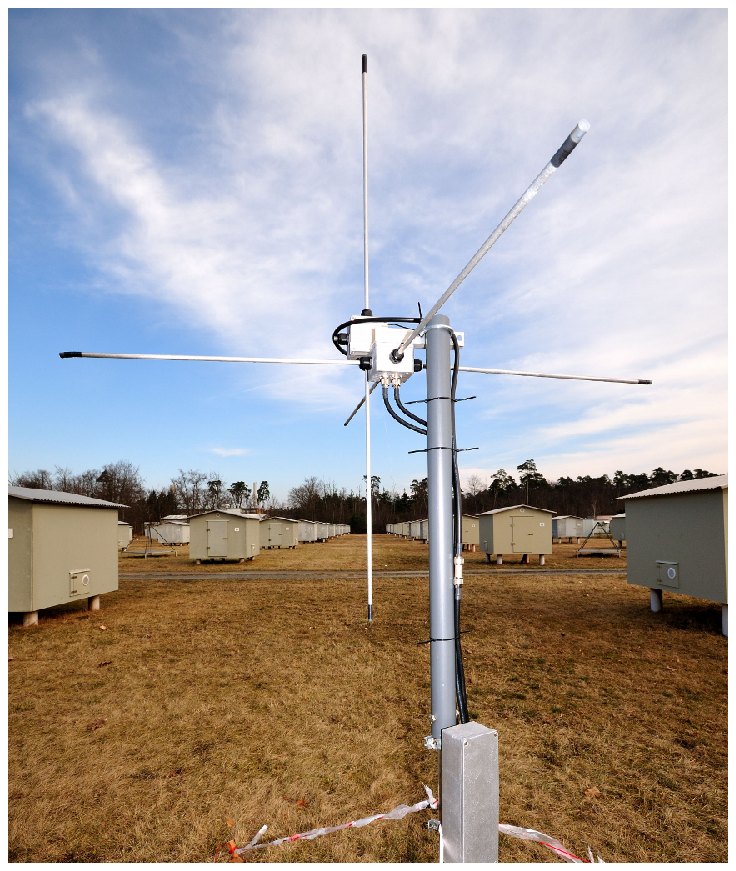}
 \label{tripole}
 \caption{Fotography of one tripole antenna as used for \mbox{LOPES-3D}}
\end{figure}

\section{Introduction}
LOPES \citep{huberlopes3d, FalckeNature2005} is an ever evolving radio antenna array co-located with the KASCADE-Grande \citep{kascade, kascade-grande} air-shower
particle-detector array at the Karlsruhe Institute of Technology (KIT). It measures the radio emission \citep{Haungsradio} from
cosmic-ray induced air showers via digital radio interferometry at energies larger than $10^{16.5}$\,eV. The latest setup of LOPES is called \mbox{LOPES-3D}. Within this last upgrade the LOPES experiment was equipped with antennas that are sensitive to all three components of the electric field vector, the so-called tripole antenna \citep{tripole}, see figure \ref{tripole}. One tripole antenna consists of three perpendicular dipole antennas and is the most straightforward approach to build an antenna sensitive to all three components of the electric field vector. With the additional vertical antenna LOPES is now able to measure the complete electric field vector directly which increases the sensitivity to inclined showers. Furthermore the comparison with emission models will improve since more information can be provided. \mbox{LOPES-3D} is reliably taking data since May 2010.

\section{calibration}
After the reconfiguration of the experiment a recalibration was necessary. The calibration procedure can be divided in several steps that can be performed independently from each other.
\begin{itemize}
\item Simulation of the antenna gain pattern using 4NEC2X \citep{4NEC2},
\item measuring the delay of the electronics using two methods and monitoring the timing using the beacon \citep{SchroederTimeCalibration2010},
\item measuring the antenna positions and
\item performing an absolute amplitude calibration \citep{NehlsHakenjosArts2007} to know which fieldstrength corresponds to which value of the analogue digital converter.
\end{itemize}
 
 \begin{figure}[!h]
 \center
 \includegraphics[height=0.5 \columnwidth ,angle = -90]{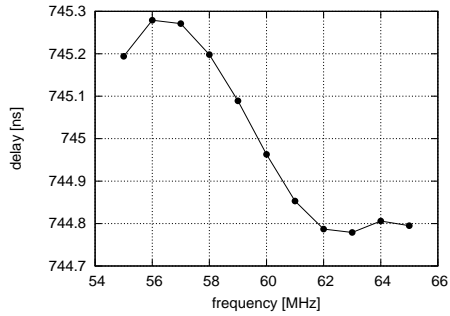}
 \label{timimnghoria2}
 \caption{Relative delay between channel 6 and channel 14 of the \mbox{LOPES-3D} experiment. Upper plot: 50 measurements for frequencies between $55$ and $65$\,MHz in $1$\,MHz steps. Lower plot: frequency dependent delay (averaged values from upper plot).}
\end{figure}
For the measurement of the delay in the electronics two methods were used. A classic one with a pulse generator connected to the electronics and the measurement of the time between sending and receiving the pulse. The newly developed method works with the phase differences of different frequencies. With the new method the frequency dependent delay of the electronics can be determined, see figure \ref{timimnghoria2}. Here the relative delay between two channels of the \mbox{LOPES-3D} experiment is shown for frequencies between 55 and 65\,MHz. The variation between different measurements is very small and far better than the required 1\,ns accuracy. This method is only limited by the intrinsic detector noise and therefore gives the best achievable result.

 \begin{figure*}[!h] \center
\includegraphics[width=0.8 \textwidth]{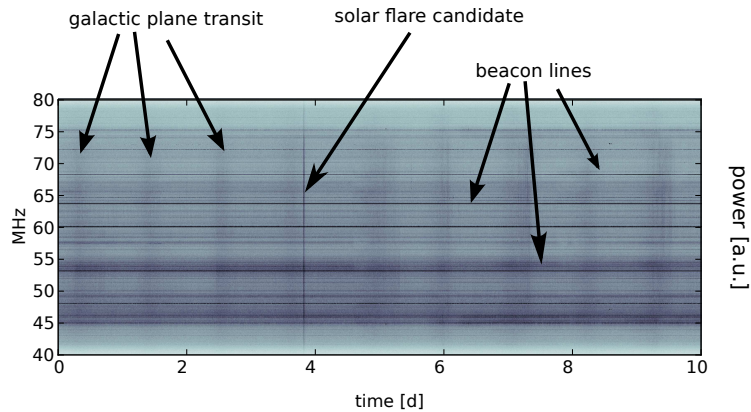}
 \label{sidnoise}
 \caption{Dynamic spectrum taken with the east-west aligned part of a \mbox{LOPES-3D} antenna. The variation in the background noise is due to the galactic plane passing by. The relatively short noisy part at the end of day 3 is most probably a solar flare.}
\end{figure*}
 
\begin{figure} \center
\includegraphics[width=.475 \textwidth]{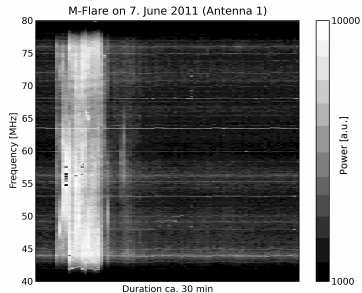}
\caption{Solar flare recorded with LOPES-3D. Power in arbitrary units.}
\label{flare3d}
 \end{figure}

\begin{figure}[!h] \center
\includegraphics[width=.9 \textwidth]{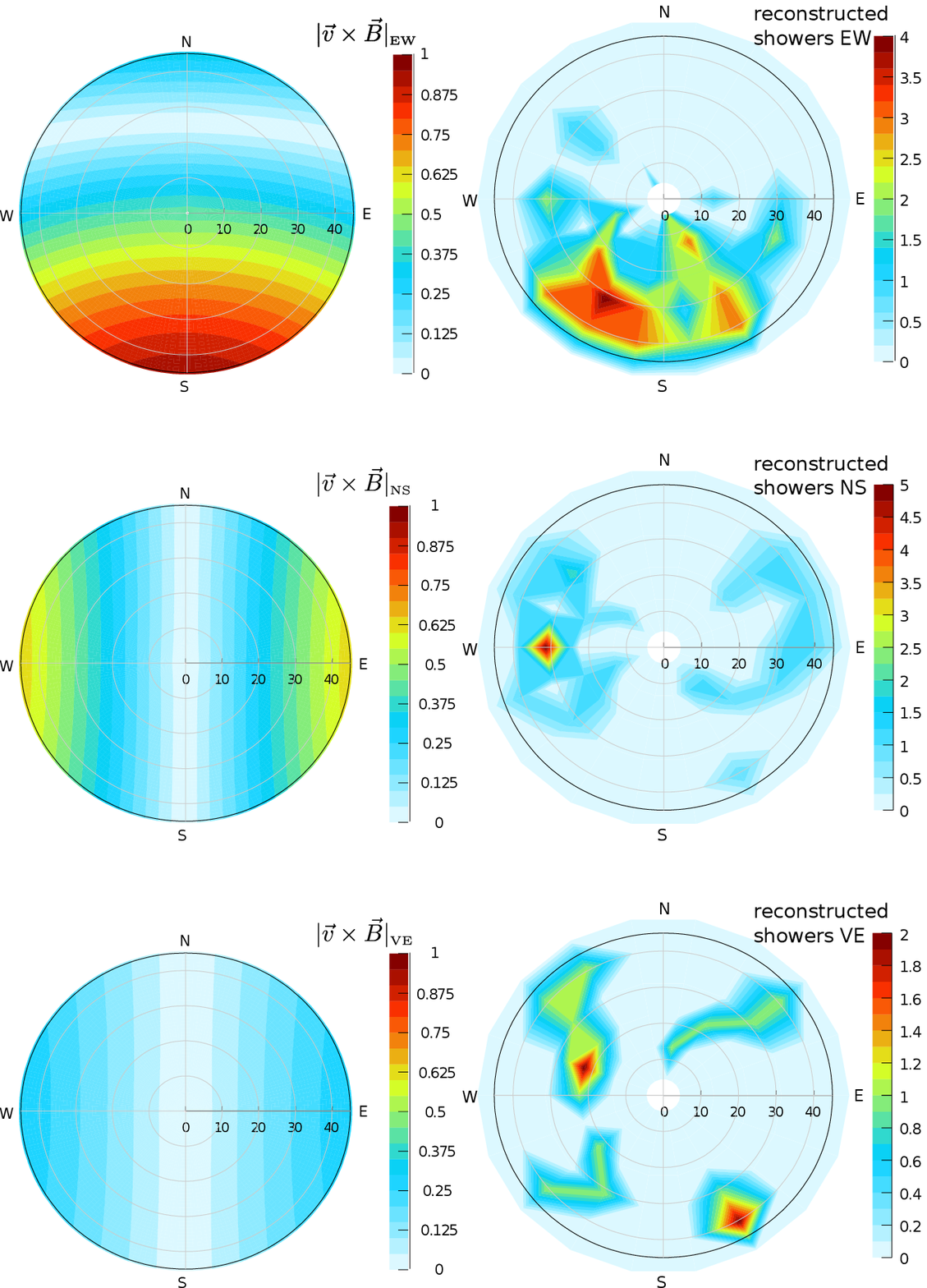} 
\caption{Comparison of reconstructed events measured with \mbox{LOPES-3D} (right hand side) to the prediction of the $\vec{v}\times\vec{B}$-model (left hand side) for the east-west (top), north-south (middle) and vertical (bottom) component of the normalized emission vector.}
\label{vxbcompare}
\end{figure}

\section{Monitoring and Sensitivity}
During the reconfiguration of LOPES to \mbox{LOPES-3D} also the monitoring was updated. A proper monitoring is crucial when running an experiment with an uptime of nearly 100\%. To monitor the state of the experiment every 20 minutes the last triggered event is briefly analyzed in the following way:
\begin{itemize}
\item A raw spectrum, i.e. the fast Fourier transform of all antenna traces is taken and plotted,
\item the average noise is calculated and shown for the last 20 days and
 \item the phase differences of the beacon signals are shown for the last 100 analyzed events.
\end{itemize}
With this monitoring, failures in electronics can be identified on a very short time scale without detailed knowledge of the experiment. In the background noise a periodicity with a $\approx24$\,h period is observed. An analysis of this periodicity indicated an origin from the galactic plane transit and therefore  the exact period is 23\,h 56\,min. In figure \ref{sidnoise} a dynamic spectrum from the data taken with an east-west aligned antenna is shown. The plane transit can be seen only with small amplitude variations. This is because the LOPES site is an industrial area with a lot of ambient noise. LOPES was designed as a prototype station and never meant to perform highly sensitive measurements. Nevertheless the observation of the galactic plane transit demonstrates that LOPES is a very sensitive radio antenna array, even with the high noise level present at the LOPES site. In the dynamic spectrum shown in figure \ref{sidnoise} a period with very high noise can be observed at day 3. This is most likely a solar flare seen by \mbox{LOPES-3D}.


\section{Performance}
When reconfiguring an experiment, the performance after the modification needs to be checked to ensure that no mistake was made during this process. In order to verify the performance of \mbox{LOPES-3D} several crosschecks were done. One possibility is to compare event rates before and after the rebuilding. For LOPES this is not straightforward since the antenna positions are reduced and this needs to be taken into account. However with two predictions: an optimistic and a pessimistic one, it is proven that LOPES fits well within the expectations, see table \ref{eventrates} and for more details reference \cite{huberlopes3d}. Another possibility, at least for radio antenna arrays, is to look at solar flares. With LOPES 30, solar flares were  observed and since the sun is becoming more active, solar flares should also be detectable with \mbox{LOPES-3D}, see figures \ref{flare3d} and \ref{sidnoise}. Here solar flares detected with the present setup \mbox{LOPES-3D} are shown.\\ The most straightforward and convincing approach to check the performance is to compare several measured events with a model that made good predictions in the past. In figure \ref{vxbcompare} a comparison of LOPES data with the $\vec{v}\times\vec{B}$-model is shown. The $\vec{v}\times\vec{B}$-model is a model that was able to describe the radio data well and was confirmed by several other radio experiments such as CODALEMA \citep{CODALEMA} and the Auger Engineering Radio Array AERA \citep{AERA}. Therefore most of the events measured with \mbox{LOPES-3D} should be well described by this model. For the comparison shown in figure \ref{vxbcompare} a normalized emission vector predicted by the $\vec{v}\times\vec{B}$-model is used and compared with arrival directions of reconstructed cosmic ray air showers. A comparison of the relative amplitude with an arrival direction can be done because the more signal is above a certain threshold in one dedicated component of the emission vector the more likely this shower can be reconstructed. For this comparison the different polarizations were analyzed separately. For the east-west component 61 events were reconstructible and survived the very strong quality cut of a signal-to-noise ratio of 8 in power. For the north-south component 26 events survived and for the vertical 16 events survived. This is expected since most of the radio emission from extensive air showers is emitted in the east-west component, second most in north-south, and the fewest in vertical. The measured event rate, the agreement of the measured data with the $\vec{v}\times\vec{B}$-model and the detected solar flare, at the exact expected time, are evidence for the functionality of LOPES-3D.

\begin{table}
\centering
\caption{Event statistics of \mbox{LOPES-3D}}
\begin{tabular}{ll}
\hline
Average rate & [events/week]\\
\hline
LOPES 30 (EW)\citep{NehlsThesis2008} & $3.5$\\
\mbox{LOPES-3D} expected (EW) & $0.39-1.17$\\
\mbox{LOPES-3D} (only EW) & $1.06$\\
\mbox{LOPES-3D} (all) & $1.75$\\
\hline
\end{tabular}
\label{eventrates}
\end{table}

\section{conclusion}
The LOPES experiment at the Karlsruhe Institute of Technology is a first generation radio interferometer \citep{NiglAA} for the detection of cosmic rays. LOPES was reconfigured several times to further develop the radio detection technique of cosmic rays. In its latest setup, \mbox{LOPES-3D}, it is measuring all three components of the electric field vector emitted by extensive air showers directly and by this LOPES once again fulfils its role as a pioneering radio detector of cosmic rays. Studies have shown that \mbox{LOPES-3D} is behaving as expected. By comparing rates of reconstructed events with the former setup of LOPES and by the agreement of the data taken with the predictions of the $\vec{v}\times\vec{B}$-model one can be confident that \mbox{LOPES-3D} measures cosmic rays. Furthermore it could be shown that the sensitivity of LOPES is good enough to see the galactic plane transit, even in the very noisy environment where LOPES is built. With direct measurement of the vertical component and the interferometrical reconstruction \mbox{LOPES-3D} is a unique experiment that will show the prospects of vectorial measurements for future applications such as AERA \citep{AERA}, by combining the measured information from all three vector components in on analysis and not analysing all components separately.

\section*{Acknowledgments}
LOPES and KASCADE-Grande have been supported by the German Federal Ministry of Education and Research. KASCADE-Grande is partly supported by the MIUR and INAF of Italy, the Polish Ministry of Science and Higher Education and by the Romanian Authority for Scientific Research UEFISCDI (PNII-IDEI grant 271/2011). This research has been supported by grant number VH-NG-413 of the Helmholtz Association.


\end{document}